\documentclass[pra,twocolumn,floatfix,superscriptaddress,showpacs]{revtex4}

\usepackage{graphicx}

\begin{document}

\title{Persistent currents in a circular array of Bose-Einstein condensates}

\date{\today}

\author{Gh.-S.\ Paraoanu}
\altaffiliation[Also at the ]{Department
        of Theoretical Physics, National Institute for Physics and Nuclear
        Engineering, PO BOX MG-6, R-76900, Bucharest, Romania.}
\affiliation{Department of Physics, University of
Jyv\"askyl\"a, P.O.Box 35 (YFL),  40014 Jyv\"askyl\"a, Finland\\ and
Department of Physics, Loomis Laboratory, 1110 W.\ Green
Street,
University of Illinois at Urbana-Champaign, Urbana IL61801, USA}

\begin{abstract}

A ring-shaped array of Bose-Einstein condensed atomic gases 
can display circular currents if the relative phase of neighboring  
condensates becomes locked to certain values. 
It is shown that, irrespective of the mechanism responsible for
generating these states, 
only a restricted set of 
currents are stable, depending on the number of  
condensates, on the interaction and tunneling energies, and on the total number of 
particles. 
Different instabilities due to quasiparticle excitations are characterized
and
possible experimental setups for testing the stability prediction are also 
discussed.

\end{abstract}


\maketitle

\section{Introduction}

The existence of flow without dissipation is a signature of superfluidity and superconductivity. This phenomenon has been studied in various 
condensed-matter systems, most notably in helium and metallic superconductors. The experimental achievement of
Bose-Einstein (BEC) condensation in alkali-metal atomic gases \cite{BEC} opened the possibility of investigating the superfluid properties of dilute BEC
gases trapped in various external potentials generated by magnetic or laser fields. The geometry of the trapping external potentials can be engineered
into various shapes: from almost spherically symmetric to highly anisotropic, from double wells to lattices.                    

We study, in the present paper,  the stability of currents for a system of $M$ condensates in a topologically restricted configuration, with tunneling allowed
only between neighboring condensates. This type of trap can be created, for instance, for 
not too large $M$'s,  by using an experimental setup similar to that of the MIT interference
experiment \cite{int} but with a spatial shape of the blue-detuned laser beam, which would ensure the separation of the gas into three or more pieces. Another possibility involves the 
use of two-dimensional optical lattices in magnetic traps of the same type as
those recently employed to study the superfluid-Mott insulator transition \cite{dieter}. Overlapping a highly repulsive optical potential (for example, generated by a blue-detuned cylindrical laser beam) in the middle of the magnetic 
trap results in the creation of a Mexican-hat potential for the atoms already confined in the lattice.

A simple application of the concept of spontaneously broken gauge symmetry to this system
has as a consequence the appearance of well-defined phase relations between consecutive
condensates and the flow of a circular current around the ring. 
Nowadays there is a  variety of experimental methods for inducing  currents into an already formed condensate:
tilting the lattice in a gravitational field \cite{alot} or accelerating it \cite{acc}, displacing  the enveloping external potential created by magnetic trapping \cite{sm}, phase imprinting \cite{ph}, coherent conversion between two hyperfine states (one serving as the pinning potential) \cite{primu}, or stirring the 
condensate with a laser beam \cite{st}.

In this paper we show  that, no matter what mechanism is chosen
to attempt to generate these circular flow states,
some of them are in fact unstable, 
either dynamically or thermodynamically. 
The states that are stable and produce persistent currents are identified as those with circulation below a certain critical value, which in general depends
on the parameters related to on-site interaction such as density and scattering length, on the tunneling rate, and on 
the number $M$ of condensates. The maximum critical circulation (vorticity) of these states is  $M/4$.

In the following we will briefly discuss the relevance of our study for the
two processes that involve at different points the concept of broken phase symmetry:
the Kibble-Zurek mechanism and the non-adiabatic quantum phase transitions.

The Kibble-Zurek mechanism \cite{kibble} predicts the appearance of topological defects in systems undergoing a rapid quench through a continuous phase transition. For the case of dilute alkali atomic gases it has been already argued that during a fast quench vortex lines will be formed \cite{ang}. 
It is likely however that a Kibble-Zurek experiment with atomic gases will not be 
realized in the way originally imagined, but rather by crafting traps 
simulating a closed array 
of condensates which can tunnel from one site to another, in a way 
similar to the experiments done using Josephson junctions \cite{raz}. This is precisely the 
system analyzed in this paper. A fast evaporative cooling of a gas of bosons in this type of trap would result in the formation of domains with broken phase symmetry 
in the wells of the potential (due to higher particle densities). The overlap
of the order parameters leads to transfer of atoms between neighboring wells. 
But in the end, the probabilities assigned to different outcome states
by any  microscopic analysis (time-dependent Ginzburg-Landau GL, kinetic theory, etc.) of the non-equilibrium 
problem have to be supplemented with the constrain that only the stable currents will
survive - indeed, as shown in this paper, modes that otherwise would pass unsuppressed
into the order parameter can decay by the emission of quasiparticles.

A similar conclusion can be drawn for the second process mentioned above, namely 
 quantum phase transitions. The superfluid - Mott insulator transition has been 
achieved and studied experimentally intensively in recent times in two-dimensional
optical lattices \cite{dieter}; the same ideas can be applied for an array of the type described in this paper. In this case crossing the 
quantum critical point results in a coherent superposition of states with different circulations \cite{j}. The mechanism that breaks the phase symmetry is then any decoherence process:
a measurement, particle losses, etc. But, roughly, as shown below, only half of the macroscopically occupied states that result in this
way will be persistent currents.

\section{A simple model}

Let us consider a simple model that captures the essential 
stability features of a macroscopically occupied mode
in the 
topologically constrained configuration described above.

We consider $M$ identical small condensates in contact with each other and 
a 
cylindrical system of coordinates $(r,\theta ,z)$.  The total number of 
atoms is $N$. The centers of the condensates 
are positioned  at $\vec{R}_{\lambda} = (R, \theta_{\lambda}, 
z=0)$, where 
$\theta_{\lambda} = 
2\pi \lambda /M$ and $\lambda$ runs from $0$ to $M-1$.

The Hamiltonian of the system is 
\begin{equation}
H = \int d\vec{r} \hat{\psi}^{\dagger} (\vec{r})\left[-\frac{\hbar^{2}}{2m}\Delta
 + V(\vec{r}) + \frac{g}{2}\hat{\psi}(\vec{r})^{\dagger}\hat{\psi}(\vec{r})\right]\hat{\psi} 
(\vec{r}),\label{unter}
\end{equation}
where $g=4\pi\hbar^{2}/m$ and $a$ is the scattering length.

The external potential $V(\vec{r})$ is crafted to be 
high enough around the origin, 
so that the atoms cannot penetrate there; it also has a number of $M$ 
minima  at $(R, \theta_{\lambda})$, around which the condensed atoms 
tend 
to localize. The delocalization effect comes from the possibility of 
tunneling between nearby wells.
We also assume that each condensate has a small enough number of 
particles 
so that they are not in the Thomas-Fermi regime (for wells of dimension $\approx 1\mu $m and the scattering length of Na or Rb, the number of particles 
in each condensate can be at most in the hundreds).  
In this situation, the
wavefunctions of each condensate depend weakly on the number of atoms 
in the well \cite{milburncorney}, and one can apply a   
$M$-mode approximation  for the field operator
\begin{equation}
\hat{\psi} (\vec{r}) = \sum_{\lambda =0}^{M-1}\phi 
(\vec{r}-\vec{R}_{\lambda}) 
\hat{a} (\lambda ),\label{tacar}
\end{equation}
where $\phi$ is a solution of the Schr\"odinger equation for each 
well;
 the Hamiltonian (\ref{unter}) takes the Bose-Hubbard form
\begin{eqnarray}
H &=& 
-t\sum_{\lambda 
=0}^{M-1}\left[\hat{a}^{\dagger} (\lambda )\hat{a} (\lambda +1 )
+ \hat{a}^{\dagger}(\lambda + 1)\hat{a}(\lambda )\right] \nonumber \\
&&+ \frac{w}{2}\sum_{\lambda 
=0}^{M-1}\hat{a}^{\dagger} (\lambda )\hat{a}^{\dagger} (\lambda )
\hat{a} (\lambda )\hat{a} (\lambda ) .\label{miltonian}
\end{eqnarray}
Here the  constant terms are omitted;  the tunneling matrix element 
 is given by
\begin{equation}
t = \int d\vec{r}\phi^{*} (\vec{r}-\vec{R}_{\lambda})\left[\frac{\hbar^{2}}{m}
\Delta - V(\vec{r})\right]\phi (\vec{r}-\vec{R}_{\lambda +1}),
\end{equation}
and the on-site energy is
\begin{equation}
w = g\int d\vec{r}|\phi (\vec{r})|^{4}.
\end{equation}
The boundary conditions are periodic by construction, $\hat{a} (\lambda 
) =
\hat{a}(\lambda + M)$.
The $M$-mode approximation and the relation (\ref{tacar}) can be seen 
also as an expansion of the field operator in the Wannier basis or a tight-binding approximation.

The Hamiltonian (\ref{miltonian}) can be diagonalized in the 
Bogoliubov 
formalism. We start first with identifying the solutions of the 
Gross-Pitaevskii equation. This can be done by seeing the $\hat{a}$'s 
as field operators and $\lambda$ as a discrete coordinate. Then the first 
term
of the Hamiltonian (\ref{miltonian}) is the equivalent of the kinetic 
energy and the second term is the interaction energy. To find the 
Gross-Pitaevskii state we expand the field operator into M modes 
\begin{equation}
\hat{a} (\lambda ) = \sum_{k}\chi_{k}(\lambda )\hat{b}_{k},
\end{equation}
where the modes $\chi_{k}(\lambda )$ are periodic
$\chi_{k}(\lambda ) = \chi_{k}(\lambda + M)$, and $k$ takes $M$ positive and negative integer values in the interval $(-M/2,M/2]$. We
minimize the mean-field energy on a state
$(N!)^{-1/2}b_{q}^{\dagger N} | 0\rangle $
with the restriction that 
the total 
number of particles is constant
\begin{equation}
\frac{\delta}{\delta \chi}\left[\langle H\rangle - \mu \sum_{\lambda 
=0}^{M}\langle \hat{a}^{\dagger}(\lambda )\hat{a}(\lambda 
)\rangle\right]
 = 0 .\label{min}\end{equation}
This gives the one-dimensional lattice Gross-Pitaevskii equation 
\begin{equation}
-t\left[ \chi_{q} (\lambda +1) + \chi_{q} (\lambda 
-1)\right]
+ Nw |\chi_{q} (\lambda )|^{2}\chi_{q} (\lambda ) = \mu 
\chi_{q} 
(\lambda ).\label{grosspit}
\end{equation}
The analogy with the usual continuous Gross-Pitaevskii equation is 
even more 
transparent if one notices that if $\chi (\lambda )$ does not vary 
much from site to site, one can define
\begin{equation}
\frac{d^{2}}{d\lambda ^{2}}\chi (\lambda )\simeq \chi(\lambda + 1)
+ \chi (\lambda -1) -2\chi (\lambda ).
\end{equation}
We can also write the Hamiltonian (\ref{miltonian}) in the 
momentum basis, defined by 
\begin{eqnarray}
\hat{b}_{k} &=& 
\frac{1}{\sqrt{M}}\sum_{\lambda 
=0}^{M-1}e^{-i(2\pi/M)k\lambda}\hat{a}(\lambda),\label{poker1}
\\
\hat{b}^{\dagger}_{k} &=&
\frac{1}{\sqrt{M}}\sum_{\lambda=0}
^{M-1}e^{i(2\pi/M)k\lambda}\hat{a}^{\dagger}
(\lambda ).\label{poker2}
\end{eqnarray}

The result is, with the indices $k$, $k'$, and $l$ taking all integer
values in the interval $(-M/2,M/2]$,
\begin{eqnarray}
H = -2t\sum_{k}\cos\left(\frac{2\pi}{M}k\right)
\hat{b}^{\dagger}_{k}\hat{b}_{k} + \frac{w}{2M}\sum_{k, k', l}
\hat{b}^{\dagger}_{k+l}\hat{b}^{\dagger}_{k'-l}\hat{b}_{k'}
\hat{b}_{k}, \nonumber
\end{eqnarray}
which is formally the Hamiltonian for a uniform system of free bosons 
with kinetic energy $
-2t\cos\left(2\pi k/M\right)$
instead of the usual 
$\hbar^{2}k^{2}/2m$.

Equation (\ref{grosspit}) has solutions 
\begin{eqnarray}
\chi_{q}(\lambda )&=& \frac{1}{\sqrt{M}}e^{i(2\pi/M)q\lambda}, \\
\mu_{q} &=& -2t\cos\frac{2\pi}{M}q + \frac{N 
}{2M}w,
\end{eqnarray}
which is precisely a circular current state with circulation quantized by $q$. 

In obtaining the Gross-Pitaevskii equation we have assumed that
there is already a phase relation established between the 
$M$ condensates, in other words the state is superfluid and not fragmented (Mott insulator)
\cite{dieter}. The coherence between adjacent sites is achieved when
 $w \ll tN/M$, which corresponds to fluctuations in the relative phase between 
neighboring sites much smaller than 1 (see the Appendix). If this
condition is satisfied, we can distinguish two limits, depending on how
the kinetic energy per particle and the interaction energy per particle 
compare to each other:  $t\gg wN/M$ defines the Rabi regime, and 
$t\ll wN/M$ defines the Josephson regime. These limits were initially introduced 
in connection with the two-well problem \cite{leggett}, where they have been shown to correspond to number-phase coherent and squeezed  
ground states but their generalization to lattices is straightforward
(see
\cite{alot,javanainen} and the Appendix). Experimentally, achieving the 
Rabi regime and at the same time preserving the validity of the tight-binding approximation can be done by tuning the scattering length in a magnetic field (Feshbach resonance) rather than decreasing the depth of the potential wells.

Let us  now turn to the problem of the excitation spectrum. 
To derive the Bogoliubov - de Gennes equations we 
start with the time-dependent Gross-Pitaevskii equation and linearize 
it for small fluctuations of the order parameter around the macroscopically occupied mode.
We obtain
\begin{widetext}
\begin{eqnarray}
E_{k}u_{k}(\lambda ) &=& -t\left[ u_{k}(\lambda +1) +
u_{k}(\lambda -1)\right] + \left[-\mu_{q} + 2Nw|\chi_{q}(\lambda 
)|^{2}\right]u_{k}(\lambda )+ Nw\chi_{q}^{2}(\lambda )
v_{k}(\lambda ), \\
-E_{k}v_{k}(\lambda ) &=& -t\left[ v_{k}(\lambda +1) + 
v_{k}(\lambda -1)\right] + \left[-\mu_{q} + 2Nw|\chi_{q}(\lambda   
)|^{2}\right]v_{k}(\lambda )+ Nw\chi_{q}^{*2}(\lambda )
u_{k}(\lambda ) .
\end{eqnarray}
\end{widetext}
The solution of this system is found as
\begin{eqnarray}
u_{k}(\lambda ) &=& \frac{1}{\sqrt{M}}e^{i(2\pi/M)(k+q)}u_{k},\\
v_{k}(\lambda ) &=& \frac{1}{\sqrt{M}}e^{i(2\pi/M)(k-q)}v_{k},
\end{eqnarray}
with normalization $|u_{k}|^{2} -|v_{k}|^{2} = 1$; the corresponding 
equations for $u_{k}$ and $v_{k}$ become
\begin{eqnarray}
\left[-2t\cos \frac{2\pi}{M}(k+q) + 
2t\cos\frac{2\pi}{M}q + \frac{N}{M}w \right]u_{k}
 &&\nonumber \\ 
+ \frac{N}{M}wv_{k} = E_{k}u_{k},&& \label{aandre}\\
\left[2t\cos 
\frac{2\pi}{M}(k-q) -
2t\cos\frac{2\pi}{M}q - \frac{N}{M}w \right]v_{k}&&
\nonumber \\
-\frac{N}{M}wu_{k}
= E_{k}v_{k}.\label{andre}
\end{eqnarray}

Solving this system gives
\begin{equation}
E_{k}^{(\pm)} = 2t\sin\frac{2\pi}{M}k\sin\frac{2\pi}{M}q 
\pm 
\sqrt{\epsilon_{k}\left(\epsilon_{k} + \frac{2N}{M}w\right)},
\end{equation}
where we have used  the notation 
\begin{equation}
\epsilon_{k} = 
2t\cos\frac{2\pi}{M}q\left(1-\cos\frac{2\pi}{M}k\right).
\end{equation}
Replacing the solution for $E_{k}$ in Eqs.  (\ref{aandre},\ref{andre}) we obtain
\begin{equation}
\left[ \epsilon_{k} \mp \sqrt{\epsilon_{k}\left(\epsilon_{k} + 
\frac{2N}{M}w\right)} + \frac{N}{M}w\right]u^{(\pm )}_{k}  
=-\frac{N}{M}wv^{(\pm )}_{k}, \label{strong}
\end{equation}
\begin{equation}
\left[ \epsilon_{k} \pm \sqrt{\epsilon_{k}\left(\epsilon_{k} +
\frac{2N}{M}w\right)} + \frac{N}{M}w\right]v^{(\pm )}_{k} =-
\frac{N}{M}wu^{(\pm)}_{k}, \label{sstrong}
\end{equation}
with the restriction imposed by the normalization condition
$|u_{k}|^{2} - |v_{k}|^{2} = 1$ which eventually will 
force us to make a choice between the two possible solutions indexed by 
$(\pm )$.
An interesting observation concerns the negative-energy eigenstates; 
assuming that we found a positive eigenenergy $E^{(\pm )}_{k}$ with 
eigenvalues
$\left(\begin{array}{c}u^{(\pm )}_{k}(\lambda )\\v^{(\pm 
)}_{k}(\lambda )\end{array}\right)$, the corresponding negative-energy 
eigenstate
is $-E^{(\pm )}_{k}$ with eigenvalues $\left(\begin{array}{c}v^{(\pm 
)*}_{k}(\lambda )\\u^{(\pm 
)*}_{k}(\lambda )\end{array}\right)$. But $-E_{k}^{(\pm )} = 
E_{-k}^{(\mp)}$  and 
\[
\left(\begin{array}{c}v^{(\pm
)*}_{k}(\lambda )\\ u^{(\pm
)*}_{k}(\lambda )\end{array}\right)= 
\left(\begin{array}{c}\frac{1}{\sqrt{M}}e^{-i\frac{2\pi}{M}(k-q)}v^{(\pm
)*}_{k}\\ \frac{1}{\sqrt{M}}
e^{-i\frac{2\pi}{M}(k+q)}u^{(\pm
)*}_{k}\end{array}\right)
=\left(\begin{array}{c}u^{(\mp )}_{-k}(\lambda )\\v^{(\mp
)}_{-k}(\lambda )\end{array}\right),
\]
where we have used $u^{(\pm )*}_{k} = v^{(\mp )}_{-k}$ and 
$v^{(\pm )*}_{k} = u^{(\mp )}_{-k}$; these last two relations can 
be proved readily from Eqs. (\ref{aandre}) and (\ref{andre}). They show that if, say, 
$(+)$ 
for a certain $k$ is a 
positive-energy eigenstate, then its corresponding negative-energy
eigenstate  is 
a 
positive-energy eigenstate of $(-)$ for 
$-k$. In other words, the state with $E_{k}^{(-)}$ represents the 
``antiparticle'' (in the spirit of Dirac's theory) of 
the state with $E_{-k}^{(+)}$.

\section{Stability analysis}

The stability of the states can be checked out by considering 
small 
perturbations of the condensate state - by small 
alterations of 
the phase and the number of particles on each site. It is useful to 
distinguish between two types of stabilities.

\subsection{Dynamical stability}

If the eigenvalues $E_{k}$ are complex, then the system is in a 
dynamically unstable state (or in a point of Lyapunov instability). 
The 
reason is that any perturbation will be exponentially magnified -
the system tends to go as fast as it can as far as possible from that 
point.
Dynamically stable states are those for which the condition
\begin{equation}
\epsilon_{k}\left(\epsilon_{k} + \frac{2N}{M}w\right) \geq 0
\end{equation}
is satisfied for any $k$. Let us first note that $\cos 2\pi q/M = 0$
does not yield properly normalized solutions, according to Eqs.
(\ref{strong}) and (\ref{sstrong}), so we will exclude from the beginning the 
states $q=\pm M/4$.
We distinguish two cases:

1$^{\rm o}$] $\epsilon_{k} > 0$ which is equivalent to $\cos\left( 2\pi 
q/M\right) 
> 0$, or  $q$ is in the interval 
$(-M/4,M/4)$. 

2$^{\rm o}$] $\epsilon_{k} < 0$ and $-\epsilon_{k}\geq 2wN/M$. 
Since the minimum value of $-\epsilon_{k}$ is reached 
when $k=\pm 1$, it follows that 
\begin{equation}
\cos\frac{2\pi}{M}q \leq -\frac{Nw}{Mt\left( 1 
-\cos 2\pi/M\right)}.
\end{equation}
This inequality implies also $wN/M \leq t\left( 1
-\cos 2\pi /M\right)$.

In conclusion, the system is in a dynamically stable state if either
condition 1$^{\rm o}$] or 2$^{\rm o}$] is satisfied. 
In the Josephson regime, only condition 1$^{\rm o}$] can be satisfied, so only the 
modes from $-M/4$ to $M/4$ are dynamically stable. In  
the Rabi regime, $Nw/Mt\ll 1$ so we can distinguish two cases:
when $M$ is of the order of unity, clearly all the modes are stable; however, when $M$
is large, of the order of $t/Nw$, there can be dynamically unstable modes located
at $|q|>M/4$; the relative number of these modes, as a fraction of the total number $M$ of modes, is extremely small, since it is limited by $\sqrt{Nw/Mt}\ll 1$.
We conclude that, in the Rabi regime,   
most of the  states 
are dynamically stable.
The dynamically unstable states can be treated in a linearized theory 
only for periods of time that are logarithmic with respect 
to the initial state \cite{garay}, since the 
quantum 
fluctuations will trigger an exponentially divergent evolution away from 
the initial state.

\subsection{Thermodynamical stability}

A point of thermodynamical instability (or energetic instability) is  
a circular flow state which is still not a local minimum of the energy 
functional; however, small perturbations do not dynamically bring 
the system far from the initial state in the absence of dissipation. 
If dissipation is introduced (or, as in our case, if the 
system suffers thermalization by collisions), the system will not stay 
arbitrarily close to the initial state but instead decay to a 
thermodynamically stable state. The name "thermodynamical instability" 
is thus justified by the fact that such a state cannot be in 
thermodynamic equilibrium.
The thermodynamically unstable states have a real excitation spectrum 
which is  characterized by 
the existence of eigenenergies with negative 
$E_{k}$ (with the 
normalization corresponding to positive eigenstates, $|u_{k}|^{2} -
|v_{k}|^{2} = 1$), and are known to produce interesting effects in optical lattices \cite{niu}. For vortices in harmonic traps
they are responsible for the 
precession of the vortex core around the condensate axis 
\cite{doilea}. This occurs because the system in a 
thermodynamically unstable vortex state can reduce its energy by 
transferring 
particles from the condensate to the negative-energy modes; for an 
un-pinned vortex for example this happens by few-particle excitations 
to the core mode \cite{rokhsar}.
Let us now study the case of thermodynamical stability in situations 1$^{\rm o}$] and 2$^{\rm o}$] 
described above.

1$^{\rm o}$] $\cos 2\pi q /M > 0$. In this case  
only $E_{k}^{(+)}$ is a valid solution. Indeed, 
from Eq. (\ref{strong}), the condition $|u_{k}| > |v_{k}|$ can be 
satisfied if 
\begin{equation}
\left|\epsilon_{k} \mp \sqrt{\epsilon_{k}\left(\epsilon_{k} + 
\frac{2N}{M}w\right)} + \frac{N}{M}w\right| < \frac{N}{M}w,
\label{wwe}\end{equation}
and clearly the solution with the lower sign cannot satisfy it. 

Let us show that the upper sign solution always satisfies this 
inequality. If the expression under the modulus in Eq. (\ref{wwe}) is 
positive,
the inequality is fulfilled trivially. If it is negative, it becomes
$
\epsilon_{k} - \sqrt{\epsilon_{k}\left(\epsilon_{k} + 
2wN/M\right)} > -2wN/M,
$
or in another form
$
\epsilon_{k}/\left[\epsilon_{k} - 
\sqrt{\epsilon_{k}\left(\epsilon_{k} +
2wN/M\right)}\right] < 1 ,
$
which is satisfied for any values of the parameters. 

The values of the Bogoliubov amplitudes $u_{k}$ and $v_{k}$ are 
obtained from Eq. (\ref{strong}) as
\begin{eqnarray}
v_{k} &=& \frac{\sqrt{\epsilon_{k} + 
2wN/M} - 
\sqrt{\epsilon_{k}}}{2\left[\epsilon_{k}\left(\epsilon_{k} + 
2wN/M\right)\right]^{1/4}} \\
u_{k} &=& \frac{\sqrt{\epsilon_{k} + 
2wN/M} +
\sqrt{\epsilon_{k}}}{2\left[\epsilon_{k}\left(\epsilon_{k} + 
2wN/M\right)\right]^{1/4}}
\end{eqnarray}

We now have to impose the condition of thermodynamical stability: 
$E_{k}^{(+)} \geq 0$ for any $k$. This condition is certainly 
satisfied when $2\pi k/M$ is in the first and second quadrants, 
but not 
necessarily when $\sin 2\pi k/M <0$. Enforcing $E_{k}^{(+)} 
\geq 0$ for $k < 0$ yields
\begin{equation}
t\left[ 1 + \cos\frac{2\pi}{M}k - 2\cos^{2}\frac{2\pi}{M}q\right] 
\leq \frac{N}{M}w\cos\frac{2\pi}{M}q .\label{astral}
\end{equation}

The maximum of the left hand side (LHS) is achieved when $k= -1$. With
some algebraic manipulations the constraint can be put in the form
\begin{equation}
2t\sin\frac{\pi}{M}(2q-1)\sin\frac{\pi}{M}(2q+1)   
\leq \frac{N}{M}w\cos\frac{2\pi}{M}q \label{uriy}.
\end{equation}
If this condition is satisfied we have a so-called persistent current 
flowing in our geometry - this current does not decay in time (say by 
thermal fluctuations) to a lower energy state. 
As it should be, the ground state (obtained for $q=0$) trivially 
satisfies this inequality
since in this case the LHS turns negative. 
The 
thermodynamically stable states with  $q\neq 0$ are local minima of the energy 
functional; in their case, it costs energy to create excitations. 
We expect persistent currents to be those that  will form 
via symmetry-breaking mechanisms, and survive enough to be 
finally detected.

2$^{\rm o}$] There are no persistent currents in this case, as one can 
readily check. Indeed, this time $\epsilon_{k}\leq 
-2wN/M$, and in order to have 
$|u_{k}|^{2}>|v_{k}|^{2}$,
we need to impose, from (\ref{strong}), the condition
\begin{equation}
0<|\epsilon_{k}|\pm\sqrt{|\epsilon_{k}|\left(|\epsilon_{k}|-
\frac{2N}{M}w\right)}<\frac{2N}{M}w.
\end{equation} 
Only the lower sign ensures the validity of these inequalities. 
The corresponding energy will be then $E_{k}^{(-)}$,
\begin{equation}
E_{k}^{(-)} = 2t\sin\frac{2\pi}{M}k\sin\frac{2\pi}{M}q - 
\sqrt{\epsilon_{k}\left( \epsilon_{k}+\frac{2N}{M}w\right)}.
\end{equation}
But now it is obvious that we cannot have $E_{k}^{(-)}\geq 0$ for all 
$k$'s, 
because if $q <0$ then the modes with 
$k>0$ will be thermodynamically unstable, and if 
$q>0$ then the modes with $k<0$
will not be stable. Thus, the only case in which we can have 
thermodynamical stability  is case 1$^{\rm o}$].

In conclusion, the number of stable states of circular currents depends, in general, on the number of 
particles $N$, on the number of condensates $M$, and on the ratio $t/w$
(which can be controlled experimentally by varying the barrier potentials
between the condensates). 
The precise form of this dependence is given by the condition
$|q|<M/4$, combined with Eq. (\ref{uriy}).
For example, according to Eq. (\ref{uriy}) there
cannot be stable currents with $M=3$ and $M=4$, $M=5,6,7,8$ give at most two stable circular currents, $M=9,10,11,12$ give at most four stable ones, 
etc.

It is instructive to see what happens in the Rabi regime, defined as
$t \gg wN/M$, and in the Josephson regime, where $tM/N\ll w 
\ll tN/M$. In the Josephson regime it is clear that Eq. (\ref{uriy}) is 
satisfied, so we have both thermodynamical and dynamical stability
for $\cos 2\pi q/M >0$. 
More interesting is what happens in the Rabi regime. 
If $M$ is of the order of unity clearly
Eq. (\ref{uriy}) cannot be fulfilled; when $M$ becomes 
of the order of $2t/Nw$, 
there can be stable modes with momenta given by
$2\pi q/M
<\sqrt{Nw/2tM + (\pi/M )^{2}}$ . They are then a very small
fraction of the total number of modes $M$ and they lead to small 
momenta of the circulating fluid. For almost all practical purposes
these modes can be assimilated with $q=0$ - in a real experiment they
would be difficult to be distinguished from the ground state.
The fact that the rest of the modes (with larger $q$) in the Rabi regime
are thermodynamically unstable can be understood physically in a simple
way. For these modes, we can simply put $w=0$ and 
search for solutions with $u_{k} = 1$ and $v_{k} = 0$. If we look at 
Eq. (\ref{strong}), it is clear that we need to take the upper sign only 
into account. The energy in this case is 
\[
E_{k} = 2 t \sin\frac{2\pi}{M}k\sin\frac{2\pi}{M}q + 2 t 
\cos\frac{2\pi}{M}q \left( 1 -\cos \frac{2\pi}{M}k\right),\nonumber
\]
which can also be written in the more relevant form
\begin{equation}
E_{k} =  2t \cos\frac{2\pi}{M}q - 2t  
\cos\frac{2\pi}{M}(k + q).
\end{equation}
This shows that $E_{k}$ is the difference between the phase-twisting 
energy per particle corresponding to a circulation  $q$ 
and the energy of an excitation around the macroscopically occupied mode. In other words, 
to create an elementary excitation  $k$ 
relative to the circulation current, we need to transfer an atom from the 
condensate to the single-particle state $k+q$.
This 
expression is positive 
for all $k$'s only if $q=0$ - so all the flow states are thermodynamically 
unstable. Instead, we have no dynamical 
instability. It is interesting to note that it 
is 
indeed the interaction that stabilizes in the end the system: a large number of persistent 
currents 
cannot exist in its absence. Thus, phase-symmetry breaking  
mechanisms, for instance, can effectively work only for systems in which there is some extra
energy to compensate for the loss of kinetic energy due to phase fluctuations.
These results are 
summarized in Table \ref{tabelul}. In the Fock regime 
($w\gg tN/M$) we cannot 
write Bogoliubov equations, so our treatment would not be valid.

\begin{table}
\center
\caption{Stability of circular currents: the Rabi and Josephson regimes. We find that a relatively large number (compared with $M$) of 
persistent currents  can 
exist only above a certain critical value of the interaction (which turns 
out to be  
in between the Rabi and Josephson regimes) and only for certain 
circulations (or angular momenta).
}
\label{tabelul}
\begin{ruledtabular}
\begin{tabular}{||l|c||c||}
REGIME & RABI & JOSEPHSON \\ \hline \hline
defined by & $t\gg wN/M$ & $tM/N\ll w\ll tN/M$ \\ \hline
dynamically stable  & most & $|q|<M/4$ \\
\hline
thermodynamically stable  & very few & $|q|<M/4$ \\

\end{tabular}
\end{ruledtabular}
\end{table}

For thermodynamically stable states (persistent currents) the quasiparticle spectrum obtained,
\begin{equation}
E_{k}^{(+)} = 2t\sin\frac{2\pi}{M}k\sin\frac{2\pi}{M}q +
\sqrt{\epsilon_{k}\left(\epsilon_{k} + \frac{2N}{M}w\right)},
\end{equation}
is different from that of the ground state, which can serve as a 
method to 
detect a circular flow state \cite{others}. Also, similar to a phenomenon called the Sagnac effect in relativistic physics, we note that 
the quasiparticles
$k$ and $-k$ propagate with different speeds around the loop, 
because they are excitations on top of a state with a flow in a 
certain direction. We have then a  
condensed-matter equivalent of the 
Sagnac effect which is related  to the 
lifting of the degeneracy of the excitations $k$, $-k$ due to the 
rotation of the condensate.

\section{Conclusions}

We have obtained the excitation spectrum of a circular flow state 
in a gas of weakly interacting 
bosons and analyzed its
dynamical and 
thermodynamical stability. Testing these stability predictions, by changing 
the tunneling, the interaction, or the number of particles is within the reach of
present-day experimental technology.

\begin{acknowledgments} 
This work was supported by the National Science Foundation grant DMR 99-86199 and 
by the Academy of Finland under 
the Finnish Centre of Excellence Programme
2000-2005 (Project No. 44875, Nuclear and 
Condensed Matter Programme at JYFL).
The author would like to acknowledge helpful discussions with S.\ Ashhab,
 C.\ Lobo, E. Mueller,  G.\ Baym,  A.\ J.\ Leggett, and useful comments on the first version of the
manuscript from J. Anglin and B. Svistunov. 
\end{acknowledgments}

\appendix*
\section{Number fluctuations}

We give here a justification of the limits that define the Rabi and Josephson regimes. As in the case of the double well \cite{leggett}, these are set respectively by the condition of Poissonian or sub-Poissonian fluctuations
of the number of atoms in each well. 

Let us look now at the ground state of the system. We will use the same notations as throughout the main part of the paper with a superscript $0$ to indicate
that they are taken at $q=0$. Since all sites are equivalent, it is enough to
look at one of them, say $\lambda =0$. If $N_0$ is the number of atoms condensed on the mode $q=0$, then $\hat{b}_{0}\simeq\sqrt{N_0}$, and from
Eqs. (10,11) one gets
\begin{eqnarray}
\hat{n}(0) \equiv \hat{a}^{\dagger}(0)\hat{a}(0) &\simeq &\frac{N_0}{M} +
\frac{\sqrt{N_0}}{M}\sum_{k\neq 0} \left(\hat{b}_{k}+\hat{b}^{\dagger}_{k}
\right) \nonumber \\
& & + \frac{1}{M}\sum_{k\neq 0, k'\neq 0} \hat{b}^{\dagger}_{k}\hat{b}_{k'}.
\end{eqnarray}
But $N_0=N-\sum_{k\neq 0} \hat{b}^{\dagger}_{k}\hat{b}_{k}$ so ignoring the smallest terms in the expansion the number of particles field operator at $\lambda =0$
can be approximated as
\begin{equation}
\hat{n}(0) \simeq\frac{N}{M} +
\frac{\sqrt{N}}{M}\sum_{k\neq 0} \left(\hat{b}_{k}+\hat{b}^{\dagger}_{k}
\right),
\end{equation}
which implies $n \equiv \langle \hat{n}(0) \rangle = N/M$. In other words, 
the splitting of the particle number operator at any point in the lattice (the density 
operator in the limit $M\gg 1$) into a mean-field value and a fluctuating field
of zero average
results naturally from the Bogoliubov theory. Now we can calculate the 
fluctuations on the ground state
\begin{equation}
\sigma_n^2 = \langle \hat{n}^{2}(0)\rangle - \langle\hat{n}(0)\rangle ^{2}
\simeq \frac{N}{M^2}\sum_{k\neq 0}\frac{\epsilon^{0}_k}{E^{0}_{k}}.
\end{equation}
Here $\epsilon_{k}^{0}=4t(\sin \pi k/M)^{2}$ and $E^{0}_{k}=\sqrt{\epsilon_{k}^{0}(
\epsilon_{k}^{0} + 2wN/M)}$. If the ``size'' $M$ of the system is large the sum
can be transformed into an integral and calculated analytically.
The result is 
\begin{equation}
\sigma_n^2 
\simeq \frac{2n}{\pi}\arctan\sqrt{\frac{2t}{nw}},
\end{equation}
the same as obtained in Ref. \cite{javanainen} for a similar Hamiltonian 
by using a phonon approach.
The parameter that controls the fluctuations of the density (and phase)
is then $t/nw$; the Rabi regime, with Poissonian fluctuations $\sigma_n\simeq\sqrt{n}$, is obtained
for $t\gg nw$, while the Josephson regime, with sub-Poissonian fluctuations $\sigma_n\simeq (8tn/\pi w)^{1/4}\ll \sqrt{n}$ is obtained in the limit $t\ll nw$. The condition $\sigma_n\gg 1$ (or, equivalently, phase fluctuations much smaller than 1) sets the limit of validity of the Bogoliubov approach: $tn\gg w$. In the Josephson regime, the
excitation spectrum is $E^{0}_{k}\simeq \sqrt{8tnw}\sin\pi k/M$.
In the Rabi regime, $E^{0}_k\simeq\epsilon^{0}_k$ for momenta
$2\pi k/M$ much larger than a critical value $\sqrt{nw/t}\ll 1$ (that is, for 
most of the modes available in the system). Below this critical value,
if there are modes available (in other words if the minimum momenta allowed by quantization $2\pi /M$ is of the order of or less than $\sqrt{nw/t}\ll 1$), the spectrum
is phonon-like, $E^{0}_{k}\simeq \sqrt{8tnw}\pi k/M$. One can regard
the Rabi regime as being essentially a noninteracting limit since the 
fraction of excitation modes that do not have a spectrum as given
by simply setting $w=0$ is negligible both in the case of $M\gg 1$ and
$M$ of the order of unity (when, since 
$\sqrt{nw/t}\ll 2\pi /M$, all the modes correspond in fact to a noninteracting spectrum).

\end{document}